\documentclass[%
reprint,
superscriptaddress,
showkeys,
prl,
amsmath,amssymb,
aps,
twocolumns,
]{revtex4-2}

\usepackage{amssymb}
\usepackage{amsmath}
\usepackage{amsthm}
\usepackage{eucal}
\usepackage{mathrsfs}
\usepackage{graphicx}
\usepackage{dcolumn}
\usepackage{hyperref}
\usepackage[usenames,dvipsnames,x11names,table]{xcolor}
\usepackage{balance}
\usepackage{dsfont}
\usepackage{tikz}
\usetikzlibrary{shapes}
\usepackage{multirow}
\usepackage{tabularx}
\usepackage{enumitem}
\usepackage{cancel,soul}
\setlist[itemize]{align=parleft,left=0pt..1em}
\usepackage{graphicx}
\usepackage{dcolumn}
\usepackage{bm}
\usepackage{tikz}
\usepackage{mathtools}
\usepackage{esdiff}
\usepackage{comment}
\usetikzlibrary{arrows.meta,positioning, shapes.geometric}

\makeatletter
\let\@internalcite\cite
\def\cite{\def\citeauthoryear##1##2{##1, ##2}\@internalcite}
\def\shortcite{\def\citeauthoryear##1##2{##2}\@internalcite}
\def\@biblabel#1{\def\citeauthoryear##1##2{##1, ##2}[#1]\hfill}
\makeatother

\begin{document}

\preprint{APS/123-QED}

\title{
When to Boost: How Dose Timing Determines the Epidemic Threshold
}
\author{Alessandro Celestini}
\affiliation{CNR, Institute for Applied Mathematics, Rome, Italy}
\author{Francesca Colaiori}
\affiliation{CNR, Institute for Complex Systems, Rome, Italy}
\affiliation{Department of Physics, Sapienza University, Rome, Italy}
\author{Stefano Guarino}
\affiliation{CNR, Institute for Applied Mathematics, Rome, Italy}
\author{Enrico Mastrostefano}
\affiliation{CNR, Institute for Applied Mathematics, Rome, Italy}
\author{Francesca Pelusi}
\affiliation{CNR, Institute for Applied Mathematics, Naples, Italy}
\author{Lena Rebecca Zastrow}
\affiliation{ISPRA, Department of the Geological Survey of Italy, Rome, Italy }


\begin{abstract}
Most vaccines require multiple doses, the first to induce recognition and antibody production and subsequent doses to boost the primary response and achieve optimal protection. We show that properly prioritizing the administration of first and second doses can shift the epidemic threshold, separating the disease--free from the endemic state and potentially preventing widespread outbreaks. Assuming homogeneous mixing, we prove that at a low vaccination rate, the best strategy is to give absolute priority to first doses. In contrast, for high vaccination rates, we propose a scheduling that outperforms a first--come first--served approach. We identify the threshold that separates these two scenarios and derive the optimal prioritization scheme and inter--dose interval. Agent--based simulations on real and synthetic contact networks validate our findings. We provide specific guidelines for effective resource allocation, showing that adjusting the timing between primer and booster significantly impacts epidemic outcomes and can determine whether the disease persists or disappears.
\end{abstract}
\maketitle

Vaccination campaigns are among the most effective public health interventions to prevent communicable diseases. 
They effectively reduce illness, prevent deaths, and provide long--term immunity against infection. 
Immunity generally wanes over time, and booster doses are often required to sustain and extend protection.
For diseases such as measles, chickenpox, pneumonia, and COVID--19, administering these extra doses is standard practice. 
Different vaccination schedules can lead to a wide range of epidemiological outcomes and, by adequately timing doses, can be optimized to enhance immunity at individual and population levels.
Optimizing such schedules became crucial during the recent SARS--CoV--2 pandemic as public health authorities had to make swift, unprecedented decisions regarding vaccine allocation. 
According to the manufacturer's recommendations, the World Health Organization~(WHO) initially recommended a 4--week interval between the first and second doses for the mRNA--BNT162b2 and mRNA--1273 vaccines. They also considered extending this interval to a maximum of 6 weeks as a pragmatic line of action in situations where vaccine distribution was severely limited~\cite{WHO-moderna2021,WHO-pfizer2021}. 
In contrast, the UK and Canada choose to prolong the inter--dose interval as a key strategy during the initial roll--out, prioritizing the delivery of the first dose to as many eligible individuals as possible.  
This strategy followed the recommendations of the Joint Committee on Vaccination and Immunisation~(JCVI), which reckoned that minimizing the vaccine--naive population would significantly reduce the risk of severe diseases and hospitalizations in the short term~\cite{JCVI2021}. 
For the ChAdOx1 nCoV--19 vaccine~(AZD1222), this decision was supported by the phase 1--3 clinical trials, which showed increased efficacy when the interval between doses was extended to 12 or more weeks, compared to less than 6 weeks~\cite{2021voyseytlSingledoseAdministrationInfluence2021a}.
The JCVI recommendation also accounted for the trade--off between the increased protection offered by a delayed second dose and the heightened risk of infection that arises as the interval between doses extends, deeming the strategy the best option given the supply constraints.

The adoption of schedules that differ from those specified in marketing authorizations is not uncommon. In fact, the optimal schedule at the individual level may only sometimes align with the best overall outcome for public health, highlighting the inherent complexity and ethical issues implicit in the choice of a specific strategy. 
Specifically, a schedule that is more beneficial for individuals could potentially compromise collective health objectives. 
Therefore, there is a challenge in balancing personal benefits with public good~\cite{2021kadirenejmDelayedSecondDose2021}. 
This balance also must account for herd--protection, mediated by the reduced transmission due to the depletion of susceptible hosts, which results from optimizing coverage at the community level~\cite{1991halloranajeDirectIndirectEffects1991}. 
The indirect effect of herd immunity is especially relevant for vulnerable groups, such as immunocompromised individuals, who cannot directly benefit from vaccination.

Mathematical modeling has long been used to provide insight into how to optimize epidemic control through pharmaceutical~\cite{2016wangprStatisticalPhysicsVaccination2016} and non--pharmaceutical~\cite{2010wallingapnasuOptimizingInfectiousDisease2010, perra2021, 2022balderramasrOptimalControlSIR2022, 2022pastor-satorrassrAdvantageSelfprotectingInterventions2022,ramasco2025} interventions.  
Vaccination strategies have been addressed in relation to timing~\cite{2024castionincReboundEpidemicControl2024}, prioritization~\cite{2009medlocksOptimizingInfluenzaVaccine2009,2021pontrellifphPrioritizingFirstDoses2021, 2022gozzipcbAnatomyFirstSix2022}, hybrid immunization~\cite{2022weincSARSCoV2AntibodyTrajectories2022, 2022goldbergnejmProtectionWaningNatural2022, 2022hallnejmProtectionSARSCoV2Covid192022, 2022cerqueira-silvatlidEffectivenessCoronaVacChAdOx12022}, contact network topology~\cite{2002pastor-satorraspreImmunizationComplexNetworks2002a}, worldwide allocation~\cite{2023gozzincEstimatingImpactCOVID192023c}, and economical costs~\cite{2011klepacpnasuSynthesizingEpidemiologicalEconomic2011}. 
The effects of varying the booster--dose timing on epidemic outcomes at the population level have been addressed in~\cite{2021moghadaspbEvaluationCOVID19Vaccination2021, 2022soutoferreirapcbAssessingBestTime2022, 2024wangrsosDeviationRecommendedSchedule2024}, where the best strategy is shown to depend on the interplay between vaccine supply and relative efficacy of the first versus second dose, with a delayed strategy being generally preferable to the recommended dosing interval in case of scarce resources. 
The impact of vaccination timing on epidemic dynamics has also been explored in Ref.~\cite{arenas2024,arenas2025}, where misaligned campaigns were shown to amplify infection peaks.
Vaccinations with pathogen dynamics and evolution are considered in Ref.~\cite{2021saad-roysEpidemiologicalEvolutionaryConsiderations2021}. 
For the SARS--CoV--2 pandemics, England's delayed second-dose vaccination strategy is validated in Ref.~\cite{imai2023quantifying}, in the context of limited resources, by comparing the epidemic outcome with the counterfactual scenario.
For a general review of statistical mechanics of vaccination see~\cite{PhysReP664_2016}.

\begin{figure}[t!]
\centering
\includegraphics[width=0.35\textwidth]{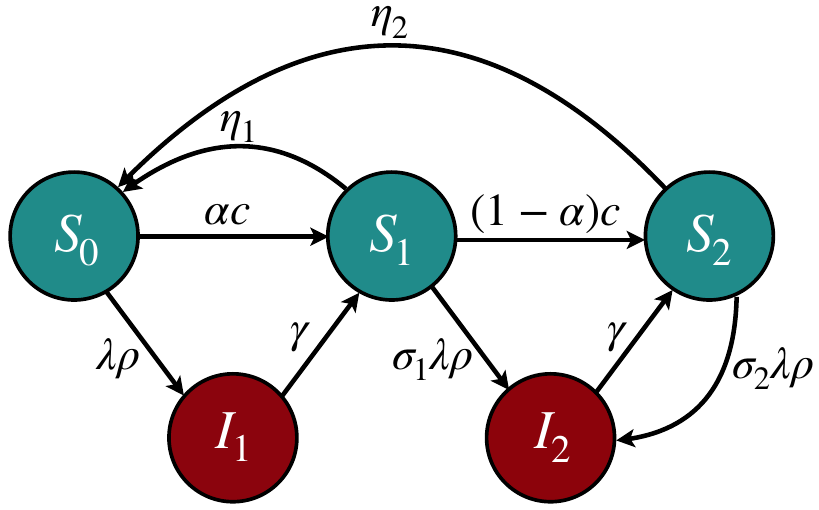}
\caption{Flow diagram of the $S^3I^2$ model with transition rates.}
\label{fig:model}
\end{figure}

Following the existing literature, we examine how booster dose timing impacts epidemic outcomes at the population level and we provide specific design guidelines for an optimal vaccination scheme that can significantly improve epidemic control.
For SARS--CoV--2 and other common infections, immunity levels can depend on vaccination status, as well as an individual’s history of infection and reinfection with the virus. 
Thus, we incorporate {\it hybrid immunization} into the model, where both types of contact provide protection.
We assume that the first immunization event -- whether induced by a vaccination shot or through natural infection -- offers short--term and partial immunity. 
A booster dose or reinfection within the waning period strengthens the protection and makes it long--lasting.
To tackle the problem, we introduce a generalized susceptible--infected--susceptible (SIS) compartmental model that includes a two--dose vaccination and accounts for both partial and waning immunity.
We will refer to this model as $S^3I^2$. 
In the $S^3I^2$ model, each individual can be in one of five states: three susceptible states ($S_0, S_1$, and $S_2$) and two infectious ones ($I_1$ and $I_2$). 
The subscript is a counter of the immunization events the individual has experienced and resets to zero after a waning time, which depends on the individual's immunity status. 
The dynamics of the model is schematized in Fig.\ref{fig:model} and proceeds as follows. 
Naive individuals in the state $S_0$ experience a reduction in susceptibility and flow to $S_1$ upon receiving their first induced or natural immunization. 
In the case of induced immunization, this transition occurs directly at rate $c\alpha$. 
Here, $\alpha\in\left[ 0,1\right]$ measures the priority given to the waiting list for the first dose compared to the booster shot, and $c$ controls the overall timescale for the vaccination process.
In the case of natural immunization, individuals in the state $S_0$ become infected and flow to $I_1$ at a rate $\lambda \rho$, where $\rho$ is the total density of the infectious population, and $\lambda$ is the transmission rate  $\lambda=\beta \langle k \rangle$, where $\beta$ is the individual transmission rate and  $\langle k \rangle$ is the average contact capacity of the nodes. From $I_1$, they then flow to $S_1$ at a rate $\gamma$ by spontaneous recovery. 
The flow between $S_1$ and $S_2$ is perfectly analogous, except that $\lambda$ is reduced by a factor $\sigma_1\leq 1$ to account for the partial immunity gained through prior exposure. 
Individuals recover at the same rate $\gamma$ independently on the counter. 
The second immunization event generally confers better protection, though not complete. As a result, individuals in the $S_2$ state can still become infected, but the transmission rate $\lambda$ is reduced by a factor $\sigma_2 \leq \sigma_1$ compared to the unprotected population.  
We generally consider finite waning times, given by the inverse of the rates $\eta_1$ and $\eta_2\leq\eta_1$, for the partial and maximal immune states, respectively. 
The interesting range for the priority index is $1/2 \leq \alpha \leq 1$, where $\alpha$ interpolates between the case where doses are administered in a first--come first--served order and the extreme where primer doses have absolute priority. 
The time interval between the two doses is $\tau= 1/(c(1-\alpha))$, and the limiting cases above correspond to $\tau=2/c$ and $\tau \rightarrow \infty$, respectively.  
The case $0 \leq \alpha \leq 1/2$ indicates a prioritization of second doses, which is of limited relevance in practical situations. 
However, for the sake of completeness, we consider $\alpha$ values in the entire $[0,1]$ interval in the following discussion. 
The dynamics is described by 
\begin{equation}
\left\{
\begin{array}{l}
\dot{s_0} = -\lambda\rho s_0-c\alpha s_0 +\eta_1 s_1 + \eta_2 s_2\\
\dot{s_1} =-\sigma_1\lambda\rho  s_1 + c\alpha s_0-c(1\! -\alpha)s_1-\eta_1 s_1 +\gamma \rho_1\\
\dot{s_2} = -\sigma_2\lambda \rho s_2+c(1-\alpha) s_1-\eta_2 s_2+ \gamma \rho_2\\
\dot{\rho_1} =\lambda \rho s_0 -\gamma \rho_1  \\
\dot{\rho_2} = \lambda\rho (\sigma_1 s_1 + \sigma_2  s_2)- \gamma \rho_2 
\end{array}
\right.
\label{eq:Dynamics}
\end{equation}
where $s_i$ and $\rho_i$ indicate the population fraction in the three susceptible and two infected states and $s_0+s_1+s_2+\rho_1+\rho_2=1$. 
The dependency on time is understood. The parameters in Eq.~\eqref{eq:Dynamics} satisfy $0\leq\sigma_2\leq\sigma_1\leq 1$, $0\leq\eta_2\leq \eta_1$, $0\leq\alpha\leq 1$. 
The first two constraints ensure the second immunization--event reinforces the effect of the first in terms of efficacy and duration. 
The third constraint fixes the range of $\alpha$ as discussed previously. 
At stationarity, the total fraction of the infectious population $\rho=\rho_1+\rho_2$ is constant, therefore,
\begin{equation}
    \rho \left(\lambda(s_0+\sigma_1 s_1 +\sigma_2 s_2) -\gamma \right)=0 \,.
    \label{rodoteq0}
\end{equation}
\begin{figure}[t!]
\centering
\includegraphics[width=.47\textwidth]{
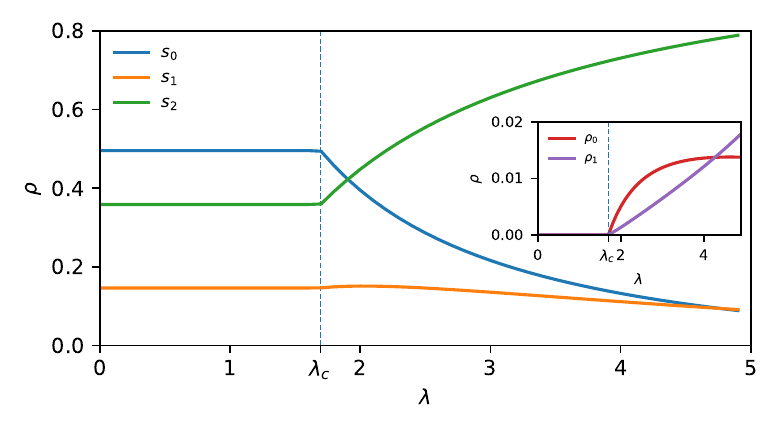}
\caption{Densities of the five subpopulations from the numerical solution of the stationary states of Eq.~\eqref{eq:Dynamics} versus $\lambda$, for $\alpha=0.6$. Other parameters were fixed to $\gamma=1$, $\sigma_1=0.4$, $\sigma_2=0.1$, $\eta_1=0.1$, $\eta_2=0.01$ and $c=0.0613$.}
\label{fig:alphastarmeanfield}
\end{figure}
\begin{figure}[t!]
\centering
\includegraphics[width=0.48\textwidth]{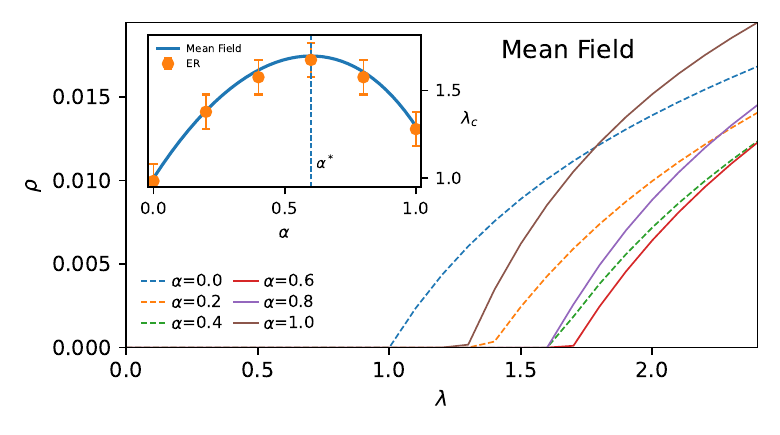}
\caption{ Mean field stationary solutions for the fraction of infectious population $\rho$ for different values of  $\alpha$. The inset shows $\lambda_c$ versus $\alpha$. The blue line is (Eq.~\ref{betac}). The symbols refer to numerical estimates of the ER
network, rescaled by $(\langle k\rangle-1)/\langle k \rangle$. }
\label{fig:meanfield_inset}
\end{figure}
Eq.~\eqref{rodoteq0} indicates that, at some $\lambda_c$, the system undergoes a dynamic transition from an absorbing, inactive phase with $\rho=0$ to a stationary endemic phase with $\rho>0$ (see Fig.~\ref{fig:alphastarmeanfield}). 
The transition is continuous for all combinations of parameters, as shown in the Supplemental Material~(SM).
In the inactive phase, the densities of susceptibles are:
\begin{equation}
\left\{
\begin{array}{l}
s_0^{(0)}=\eta_2(c(1-\alpha)+\eta_1)/N\\
s_1^{(0)}=\eta_2 c\alpha /N\\
s_2^{(0)}=c^2 \alpha (1-\alpha)/N
\end{array}
\right.
\label{InactivePhase}
\end{equation}
where $N=\eta_2(c+\eta_1)+c^2\alpha(1-\alpha)$.
When first doses have absolute priority ($\alpha=1$), then
$(s_0^{(0)},s_1^{(0)},s_2^{(0)})=(\frac{\eta_1}{c+\eta_1},\frac{c}{c+\eta_1},0)$: no virus is circulating and, since maximally immunized individuals eventually lose immunity, the population is either naive ($S_0$) or partially immunized with the first dose ($S_1$).  
In the limit $\eta_2 \rightarrow 0$, that is, when the second dose confers permanent immunity, the entire population is eventually protected ($s_2{(0)}=1$). 
Stability analysis can be performed linearizing Eqs.~\eqref{eq:Dynamics} around the solution~\eqref{InactivePhase}. 
The eigenvalues of the Jacobian matrix are given by $\Lambda^{(0)}=0$, $\Lambda_1=-\gamma<0$, $\Lambda^{(2,3)}= -(c+\eta_1+\eta_2)/2\pm\sqrt{(c+\eta_1-\eta_2)^2/4-c^2\alpha(1-\alpha)}$ (both real and negative), and $\Lambda_4=\gamma(\lambda-\lambda_c)/\lambda_c$, with $\lambda_c\!=\gamma/(s_0^{(0)}\!+\sigma_1 s_1^{(0)}\!+\sigma_2 s_2^{(0)})$. 
Therefore, the inactive fixed point is stable and reached asymptotically without oscillations for $\lambda \leq \lambda_c$, while it becomes unstable for $\lambda>\lambda_c$. The eigenvalue $\Lambda^{(0)}=0$ reflects the redundancy of Eqs.~\eqref{eq:Dynamics} and does not affect stability. Inserting Eqs.~\eqref{InactivePhase} in $\lambda_c$:
\begin{equation}
    \!\!\lambda_c(\alpha)=
    \frac{\gamma(\eta_2(c+\eta_1)+c^2\alpha(1-\alpha))}
    {\eta_2(c+\eta_1)-c \eta_2(1-\sigma_1)\alpha+\sigma_2 c^2 \alpha (1-\alpha)} .
    \label{betac}
\end{equation}
The optimal priority scheme is obtained with the value $\alpha = \alpha^*$ that maximizes $\lambda_c$: 
\begin{equation}
    \alpha^*=\{\alpha \in \left[0,1\right] \ | \ \lambda_c(\alpha)=\lambda^*\dot{=}\max_{\alpha\in \left[0,1\right]}\lambda_c(\alpha)\}\,.
\end{equation}
The optimal $\alpha$ has to satisfy $\alpha^2-2k_0 \alpha+k_0 +k_1 =0$,
where $k_0=\frac{1-\sigma_2}{1-\sigma_1}\frac{c+\eta_1}{c}\geq1$, and $k_1=\frac{\eta_2(c+\eta_1)}{c^2}\geq 0$. The equal signs are valid when both $\sigma_1=\sigma_2$ and $\eta_1=0$, and when $\eta_2=0$, respectively.
Notably, the optimal partition depends on the large set parameters of the $S^3I^2$ model only through the two combinations $k_0$ and $k_1$. 
This equation has a real solution in the physical region $\alpha \in [0,1]$ for $k_1 \leq k_0-1$ at $\alpha^-=k_0-\sqrt{k_0^2-k_0-k_1}$ that corresponds to a maximum of $\lambda_c$. For $k_1\geq k_0-1$, $\lambda_c$ increases monotonically in $[0,1]$ so that the maximum is realized in $\alpha=1$. 
\begin{figure}[t]
\centering
\includegraphics[width=0.42\textwidth]{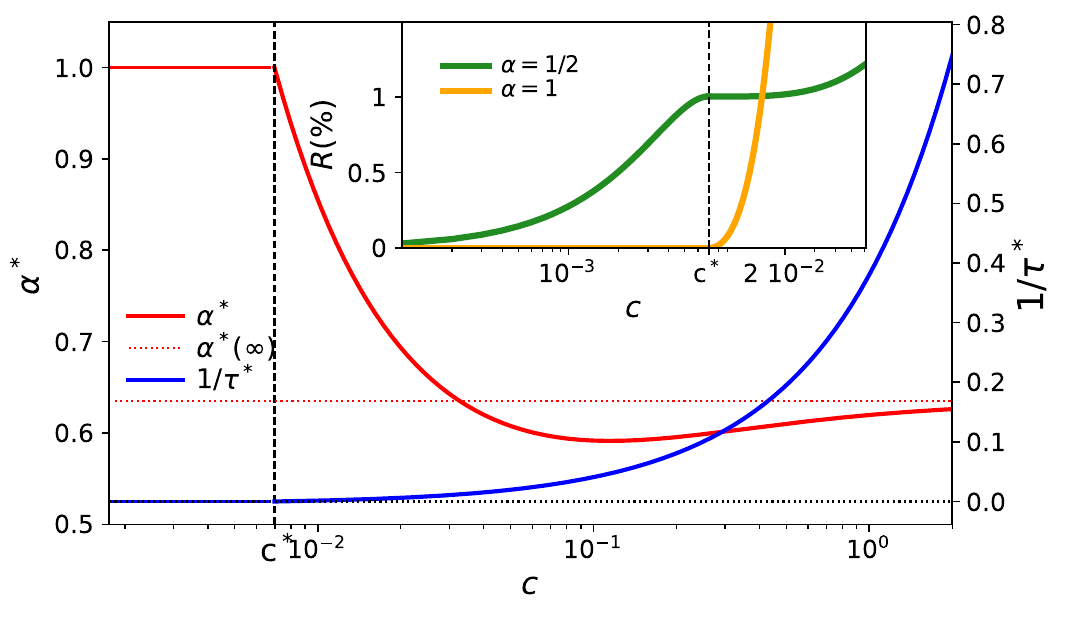}
\caption{Optimal priority index $\alpha^*$(red) and inverse inter--time $1/\tau^*$(blue) versus $c$ (semi--log). 
Inset:  Plot of the gain $R=(\lambda_c(\alpha)-\lambda_c^*)/\lambda_c(\alpha)$ between $\lambda_c^*$ resulting from the optimal prioritization schedule and $\lambda_c(\alpha)$ obtained with first--come first--served ($\alpha=1/2$, green) and the case of first doses prioritization ($\alpha=1$, yellow). The parameters values are $\eta_1 = 0.10$, $\eta_2 = 0.01$, $\sigma_1  = 0.40$, $\sigma_2  = 0.10$, and $\gamma = 1.00$. 
} 
\label{fig:AlphastarVSc}
\end{figure}
Fig.~\ref{fig:meanfield_inset} shows the mean field solutions for the stationary fraction of the infectious population $\rho$ for different values of $\alpha$.
As $\alpha$ increases, the inactive phase expands initially and then is reduced. This is clearly seen in the inset, where $\lambda_c$ is plotted as a function of $\alpha$, and shows a maximum at the predicted value $\alpha^*$. 
The symbols refer to the numerical values obtained for the Erdős--Rényi--Gilbert~(ER) graph~\cite{1959gilbertamsRandomGraphs1959a} for comparison (see below). 
The condition $k_1 \leq k_0-1$ translates into a condition on the vaccination rate $c$: there exists a critical $c^*$ such that when $c\leq c^*$, prioritizing first doses is the best choice; 
for $ c>c^*$, instead, a non--trivial priority schedule is optimal. 
The threshold value is 
$  c^*=\eta_1 (\sqrt{\Delta}-(1 -  \Sigma\eta_2/\eta_1))/(2(1-\Sigma))>0  $,
where $\Delta=((1  -  \Sigma\eta_2/\eta_1)^2+4 (1-\Sigma) \Sigma\eta_2/\eta_1)>0$, and $\Sigma=\frac{1-\sigma_1}{1-\sigma_2}\leq1$ is the relative immunity gain (the equal being valid for $\sigma_1=\sigma_2$, i.e., when the booster prolongs protection without enhancement).
At $c=c^*$, a continuous transition occurs, as shown in Fig.~\ref{fig:AlphastarVSc} where $\alpha^*$ and the optimal inverse dose--boost interval $1/\tau^*$ are plotted as functions of $c$. 
In the inset, we show the gain $R=(\lambda_c(\alpha)-\lambda_c^*)/\lambda_c(\alpha)$ between $\lambda_c^*$ resulting from the optimal prioritization schedule and $\lambda_c(\alpha)$ obtained with first--come first--served ($\alpha=1/2$, green) and with first doses prioritization ($\alpha=1$, yellow).
In summary, the best vaccination schedule corresponds to $\alpha^*=\min(\alpha^-,1)$, which equals one below $c^*$, and results in an epidemic threshold $\lambda^*_c=\lambda_c(\alpha^*)$. 
We observe that the optimal $\lambda^*_c=\gamma/\tilde{s}^{(0)}$ is obtained when $\tilde{s} = s_0 +\sigma_1 s_1+ \sigma_2 s_2$, which plays the role of an effective density of the susceptible population, is minimal.  

\begin{figure}[t!]
\centering
\includegraphics[width=0.45\textwidth]{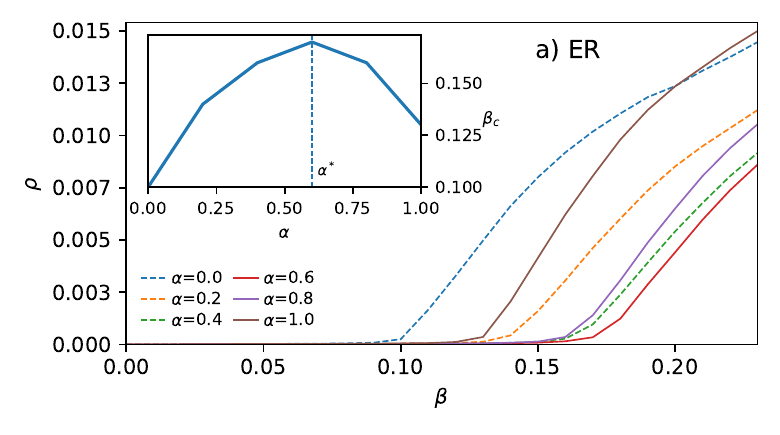}
\includegraphics[width=0.45\textwidth]{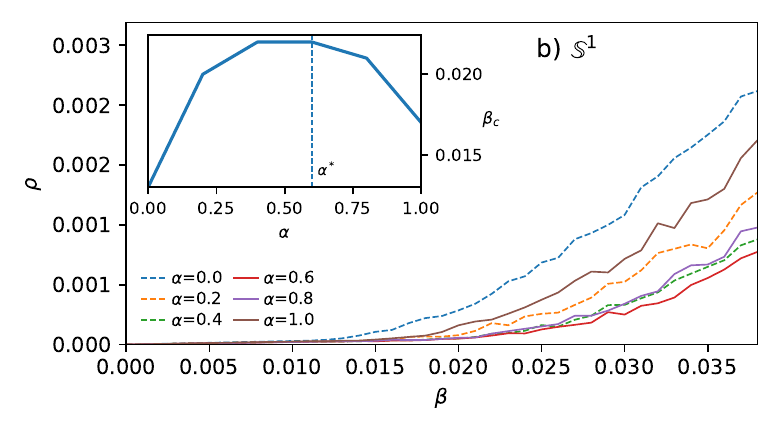}
\includegraphics[width=0.45\textwidth]{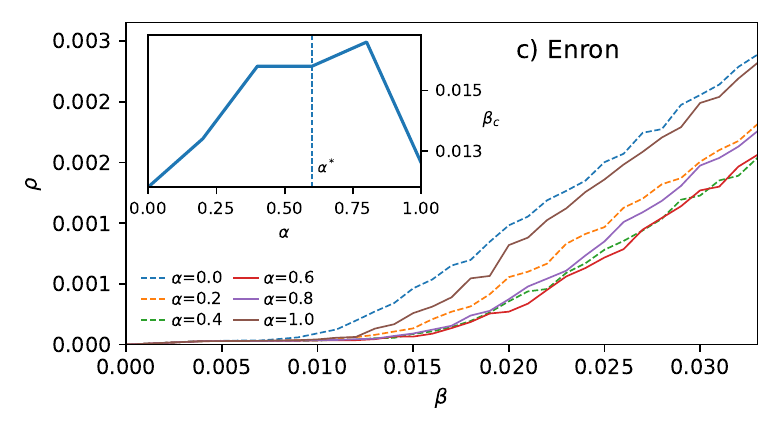}
\caption{Density $\rho$ of infectious population versus  $\beta$, for different values of $\alpha$, averaged over 50 realizations. The parameters are fixed at $\gamma=1$, $\sigma_1=0.4$, $\sigma_2=0.1$, $\eta_1=0.1$, $\eta_2=0.01$ and $c=0.0613$. The parameter $c$ has been set to have $\alpha^*=0.6$. }
\label{fig:alphastar}
\end{figure}

We simulated the $S^3I^2$ model adapting the package EoN~{\cite{eon_book,Miller2019}} on three different networks: i)  an ER graph,
ii) the Enron email network \cite{snapnets,klimt2004introducing}, and iii) a geometric random graph generated using the $\mathbb{S}^1$ model\cite{serrano2008self}. The structural properties of the networks used are summarized in Table~I of the SM.
A common issue when simulating SIS--type dynamics in finite--size systems is sampling the quasi--stationary state~\cite{PhysRevE.94.042308,COSTA2021108046}.
We implemented a method in which a subset of nodes, selected in proportion to their past activity, is reactivated whenever the dynamics get trapped in the absorbing phase~\cite{COSTA2021108046}. 
Further details are given in the SM. 
We computed the density $\rho$ of infectious individuals as the infection rate $\beta$ varies for different choices of the priority index $\alpha$ (see Fig.~\ref{fig:alphastar}). 
Following~\cite{shu2015numerical}, we estimate the critical infection rate $\beta_{c}$ for various $\alpha$. 
As shown in the insets of Fig.~\ref{fig:alphastar}, $\beta_{c}$ shows a maximum at some $\alpha_{sim}^*<1$, which confirms the theoretical predictions. 

Public health agencies need to estimate the epidemiological impact of vaccination campaigns to inform policy responses. 
A central question is how to optimize the available resources in order to minimize disease burden and mortality. 
For vaccines that provide waning immunity, adjusting the timing of booster doses significantly impacts long--term disease dynamics. 
Our study provides a solid framework to analyze this problem and indications for designing optimal vaccination schemes for endemic and emerging pathogens.
We find that, when vaccination rates are 
high, a specific prioritization scheme between the first and second doses optimizes the outcome. 
Conversely, when waiting times are long, the most effective strategy is to maximize the coverage of the first dose. 
We give an explicit expression for the threshold that separates these two scenarios and provide the exact optimal priority index for any rate above this threshold.
Tuning the inter--dose interval affects the epidemiological outcome by pushing forward the epidemic threshold and shifting the boundaries of the endemic phase. 
These findings indicate how strategically prioritizing vaccine administration can effectively suppress outbreaks that could arise under a suboptimal schedule.

Future research should address several limitations of this study. 
Firstly, the assumption of a homogeneous population may not be appropriate in most cases, as it may be more beneficial to prioritize specific population segments: vulnerable groups -- such as older adults, or key transmitters -- such as children.
The optimal strategy may differ in these scenarios, with shorter inter--dose intervals and low--coverage strategies potentially becoming more favorable. 
Secondly, a successful distribution scheme must rely on vaccine uptake; therefore, including vaccine hesitancy in the model is essential.
Lastly, we must consider the risk that strategies requiring longer to cover the population could lead to the emergence of vaccine-resistant strains, especially in the context of rapidly mutating viruses.

The $S^3I^2$ model can be generalized to include reduced infectivity after vaccination, inter--individual variability in vaccine responsiveness, asymmetry between natural and induced immunization~\cite{2022bucknercIntervalPriorSARSCoV22022}, and population replacement. \\

FC thanks Claudio Castellano and Alessandro Flammini for carefully reading the final version of this manuscript and providing valuable comments. FP acknowledges the support of the National Center for HPC, Big Data and Quantum Computing, Project CN\_00000013 -- CUP B93C22000620006, Mission 4 Component 2 Investment 1.4, funded by the European Union -- NextGenerationEU. FC and SG acknowledge support from the project “CODE – Coupling Opinion Dynamics with Epidemics,” funded under PNRR Mission 4 “Education and Research” - Component C2 - Investment 1.1 - Next Generation EU “Fund for National Research Program and Projects of Significant National Interest” PRIN 2022 PNRR, grant code P2022AKRZ9, CUP B53D23026080001.  

\bibliographystyle{apsrev4-2}
\bibliography{apssamp}

\end{document}